\documentstyle[ApJ,times,PSfig]{article}

\def\beq{\begin{equation}} \def\eeq{\end{equation}}
\def\bdm{\begin{displaymath}} \def\edm{\end{displaymath}}
 \def\eg{e.g$.$~}   \def\etal{et al$.$~}
\def\epsel{\varepsilon_e} \def\epsmag{\varepsilon_B} \def\cm3{\;{\rm cm^{-3}}} \def\deg{^{\rm o}}
\def\epse1{\varepsilon_{e,-1}} \def\epsB2{\varepsilon_{B,-2}} \def\E053{{\cal E}_{0,53}}
\def\simg{\mathrel{%
      \rlap{\raise 0.511ex \hbox{$>$}}{\lower 0.511ex \hbox{$\sim$}}}}
\def\siml{\mathrel{%
      \rlap{\raise 0.511ex \hbox{$<$}}{\lower 0.511ex \hbox{$\sim$}}}}

\begin{document}

\title{Fundamental Physical Parameters of Collimated Gamma-Ray Burst Afterglows}

\author{A. Panaitescu}
\affil{Dept. of Astrophysical Sciences, Princeton University, Princeton, NJ 08544}
\and
\author{P. Kumar}
\affil{Institute for Advanced Study, Olden Lane, Princeton, NJ 08540}

\begin{abstract}

 We determine the basic physical characteristics of eight Gamma-Ray Bursts (GRB) --
980519, 990123, 990510, 991028, 991216, 000301c, 000926 and 010222 --  by modelling the 
broadband emission of their afterglows. We find that the burst kinetic energies after the 
GRB phase are well clustered around a mean value of $3 \times 10^{50}$ ergs. In contrast, 
the energy release in $\gamma$-rays, after correcting for collimated explosion, varies 
among bursts by more than an order of magnitude.
 The jet initial apertures are the $2\deg-14\deg$ range, mildly correlated with the energy,
half of the jets being narrower than $\sim 3\deg$. This implies that, within 100 Mpc, there
are about 10 GRB remnants (expanding at $\sim 0.1\,c$) which can be resolved with VLBA.
 For all eight afterglows the total energy in the shock-accelerated electrons is close to 
equipartition with protons. However the slope of the power-law electron distribution is not 
universal, varying between 1.4 and 2.8. 
 In at least half of the cases, the density structure of the medium is inconsistent with 
an $r^{-2}$ profile. A homogeneous medium with density in the $0.1-50\cm3$ range can 
accommodate the broadband emission of all afterglows, with the exception of 990123, for 
which we find the density to be less than $10^{-2}\cm3$. If GRBs arise from the core 
collapse of massive stars, then such low densities indicate the existence of superbubbles
created by the supernovae and winds within a cluster of massive stars.

\end{abstract}

%\keywords{gamma-rays: bursts - ISM: jets and outflows - methods: numerical -
%           radiation mechanisms: non-thermal - shock waves}

\section{Introduction}

 The detection of very energetic photons (in the GeV range) in several Gamma-Ray Bursts (GRBs)
and the very short variability timescale (Fishman \& Meegan 1995), sometimes less than 1 ms, 
exhibited by the 100 keV emission of many bursts have lead to the conclusion that they arise 
from sources that are moving at extremely relativistic speeds, with Lorentz factors $\Gamma$ 
that could exceed one hundred (Fenimore, Epstein \& Ho 1993, Piran 1999). Such a highly 
relativistic motion would follow after the release of a large amount of energy in a region of
small baryonic mass. The non-uniformity in the velocity at which various parts of the outflow 
move leads to internal shocks. In this way a fraction of the kinetic energy of the outflow is 
dissipated and radiated away in $\gamma$-rays (Rees \& M\'esz\'aros 1994). Some of the remaining 
kinetic energy is converted to electromagnetic radiation when the GRB remnant is decelerated by 
the circum-burst medium (Rees \& M\'esz\'aros 1992). The resulting external shock energizes the 
swept-up gas which, similar to the GRB phase, emits synchrotron and inverse Compton emission, 
producing an afterglow. Theoretically, the fall-off of the afterglow Lorentz factor $\Gamma$ 
is expected to be a power-law in the observer time, which leads to a power-law decay of the 
afterglow flux (M\'esz\'aros \& Rees 1997). This behaviour has indeed been observed in many 
afterglows (\eg Piran 1999, Piro 2000, Wheeler 2000), on timescales of days.

 Due to the relativistic beaming of the emission, the observer receives radiation mostly from
the fluid moving within an angle $1/\Gamma$ radians off the observer's line of sight toward the
fireball center. Thus as $\Gamma$ decreases, the size of the "visible" region increases. If the 
GRB remnant is collimated into a jet, then at some time $t_j$ the entire jet surface becomes 
visible to the observer. This time is given by 
\beq
 t_j = 0.4\;(z+1) \left( E_{0,50} n_0^{-1} \right)^{1/3} \theta_{0,-1}^2 \;{\rm day} \;,
\label{tjet}
\eeq
$z$ being the burst redshift, $E_{0,50}$ the initial jet energy measured in $10^{50}$ erg,
$\theta_{0,-1}$ its initial half-opening measured in 0.1 radians, and $n_0$ the external medium 
density in $\cm3$. The lack of emitting fluid outside the jet opening leads to a faster afterglow 
decay after $t_j$. Furthermore, around $t_j$ the jet starts to expand laterally (Rhoads 1999), 
its sweeping area increases faster than before, leading to a stronger deceleration and increasing 
even more the afterglow dimming rate. The most important signature produced at $t_j$ by the jet 
collimation is the achromaticity of the afterglow emission break, which distinguishes it from the 
chromatic light-curve steepening that the passage of a spectral break through the observing band 
would yield. So far, there are eight well observed GRB afterglows (980519, 990123, 990510, 991208, 
991216, 000301c, 000926, and 010222) for which a break or a steep decay has been observed in their 
optical light-curves. These are the afterglows we model in this paper.

 The afterglow flux at a given frequency and time depends on the jet speed, on the properties of 
the external medium, and on the micro-physics of relativistic shocks. Our aim is to constrain
some of these properties by modelling the afterglow dynamics, calculating its emission and using 
the observational data to determine $i)$ the jet energy and collimation, $ii)$ the efficiency 
at which relativistic shock accelerate electrons and generate magnetic fields, and $iii)$ the 
external medium type and density.

\section{The Afterglow Model}

 There are some simplifying assumptions made in our numerical modelling, which allow the high 
computational speed necessary for a parameter space search. The most important simplification is 
that the jet front is homogeneous: the energy per solid angle and Lorentz factor have the same 
value in any direction within the jet aperture, and the internal energy density of the shocked 
fluid is uniform, at the value set by the shock jump conditions. Implicitly, we also assume that 
the jet has a sharp edge. 

 The dynamics of the jet is calculated by tracking its energy (some of which is lost radiatively),
mass and aperture, which increase as the jet expands and sweeps-up the surrounding gas (Kumar \& 
Panaitescu 2000, Panaitescu \& Kumar 2000). There are only three parameters that give the jet 
dynamics -- the initial jet energy $E_0$, initial half-angle $\theta_0$, and external particle 
density, $n$ for a homogeneous medium or the constant $A$ for a wind-like medium with profile 
$n(r) = A r^{-2}$. 

 The calculation of the afterglow synchrotron and inverse Compton emission is based on the 
assumption that the electron distribution ${\cal N}_e (\gamma)$ produced by the shock acceleration 
in the downstream region is a power-law ${\cal N}_e \propto \gamma^{-p}$ (this is supported by the 
observed power-law decay of many afterglows), starting from a minimum electron energy $\gamma_i 
m_e c^2$ and ending at a high energy break $\gamma_* m_e c^2$. The existence of this high energy 
break is due to the escape of particles from the acceleration region, to radiative losses, and is
also required to have a finite energy in electrons if $p < 2$. For simplicity we approximate the 
cut-off at $\gamma_*$ as a steeper power-law of index $q$.

 There are three or five model parameters pertaining to the microphysics of shock acceleration 
and magnetic field generation, based on which we calculate the co-moving frame afterglow spectrum.
The magnetic field strength is simply parameterized by the fraction $\epsmag$ of the post-shock 
energy density that resides in it. The distribution of the injected electrons is defined by the 
fractional energy $\epsel$ in electrons if they all had the same $\gamma = \gamma_i$ (thus $\epsel$ 
parameterizes $\gamma_i$) and  the index $p$ of the power-law distribution above $\gamma_i$.
The cut-off $\gamma_*$ is set by the the fractional energy $\epsilon$ of the electrons between 
$\gamma_i$ and $\gamma_*$.

 The synchrotron spectrum is piece-wise power-law, with breaks at the self-absorption frequency 
$\nu_a$, injection frequency $\nu_i$ corresponding to the minimum electron $\gamma_i$, cooling 
frequency $\nu_c$ corresponding to an electron cooling timescale equal to the dynamical time, 
and cut-off frequency $\nu_*$ associated with $\gamma_*$. The afterglow spectrum and light-curve
is determined by the evolution of these spectral breaks and of the flux at peak frequency 
$\nu_p = \min (\nu_c, \nu_i)$. We note that for a spreading jet ($t > t_{jet}$), the afterglow 
emission at frequencies above $\nu_i$ decays approximately as $F_\nu \propto t^{-p}$, irrespective 
of the location of $\nu_c$, provided that the electron cooling is dominated by synchrotron 
emission. Therefore the temporal slope of the post jet-break optical or radio light-curves 
gives a direct measurement of the electron distribution index $p$.
 
 We calculate numerically the afterglow emission for an observer lying on the jet axis, by 
integrating the synchrotron and inverse Compton emissions over the jet dynamics, taking into
account the variation of the bulk Lorentz factor $\Gamma$ on the equal arrival time surface.
We note that the afterglow light-curve has a weak dependence on observer location offsets less
than $\theta_0$ . The equations we employ for the calculation of the jet dynamics and of the 
received radiation (Kumar \& Panaitescu 2000, Panaitescu \& Kumar 2000, 2001) are valid in any 
relativistic regime. For the parameters we find for various GRB afterglows, the non-relativistic 
regime should set in after the last available observations. It is the mildly relativistic regime 
that corresponds to observations made later than about 10 days after the GRB event, for which a 
numerical treatment is most appropriate.

\vspace*{5mm}
\section{Features of GRB Jets}

 Based on the afterglow model outlined above, we model the broadband, time-dependent emission
of eight GRB afterglows: 980519, 990123, 990510, 991208, 991216, 000301c, 000926 and 010222, 
whose light-curves exhibited a break, offering thus the possibility of determining the initial 
jet opening $\theta_0$ and energy $E_0$.\footnote{
    Other afterglows did not have such a break for the duration spanned by the observations. 
    Although they may have been collimated, the initial jet angle cannot be determined without 
    having an observational constraint on the jet break time $t_j$. }
The basic afterglow parameters: $E_0$, $\theta_0$, $n$, $\epsel$, $\epsmag$, and $p$ (as well 
as $\epsilon$ and $q$ when they are relevant) are determined by minimization of $\chi^2$ 
(maximization of the likelihood to obtain the observed fluxes). Figure 1 shows the data and 
the model light-curves in three distinct domains: radio, optical and $X$-ray. The averages and 
dispersions of parameters for our eight GRBs, calculated from the best fit parameters of 
individual bursts, are
\bdm
 \left. \begin{array}{ll}
  E_0 = (2.6\pm 1.1)\times 10^{50} \;\; {\rm erg} \;, & \theta_0 = 6.1\deg \pm 4.5\deg \;,\\ 
  \log (n/{\rm cm^{-3}}) = 0.13 \pm 1.35    \;,       & \epsel = 0.062 \pm 0.045    \;,   \\ 
  \log \epsmag = -2.4 \pm 1.2               \;,       &  p = 1.87 \pm 0.51          \;.   
  \end{array} \right. \; 
\edm

 Figure 2 shows the best fit parameters and their 90\% confidence levels, determined by variation 
of $\chi^2$ around its minimum, for our set of eight afterglows. As illustrated, we find that jet 
energies that are clustered between $10^{50}$ and $\sim 6\times 10^{50}$ erg,\footnote{
  Using the analytical expression for the jet break time, Frail \etal (2000) have calculated 
  jet apertures $\theta_0$ for more afterglows and found that the energies released in 
  $\gamma$-rays by GRB jets are also clustered within a factor of 4.}
somewhat smaller than the typical kinetic energy of a supernova. Thus only a very small fraction 
($\sim 10^{-4}$) of the energy budget expected for GRB progenitors (Paczy\'nski 1998, M\'esz\'aros, 
Rees \& Wijers 1999, MacFadyen, Woosley \& Heger 2001) is given to highly relativistic ejecta. 
It is then surprising that the variation of this fraction among bursts is merely a factor of a few.

 We have found that models with a homogeneous medium can explain well the broadband emission of 
all eight afterglows, although in one case (000926) the best fit is unsatisfactory ($\chi^2$ is 
rather large). For 980519 and 990510 we determine external particle densities around $\sim 0.1\cm3$, 
typical for warm interstellar medium, while for 991208, 991216, 000301c, 000926 and 010222 the 
densities we find are around $10 \cm3$. For 990123 we obtain an external density below $10^{-2} 
\cm3$, similar to the afterglow 980703 (Panaitescu \& Kumar 2001).

 Our afterglow calculations also show that an $r^{-2}$ density profile of the external medium is 
compatible with the emission of 991208 and 991216, and can marginally accommodate the afterglows 
000301c and 010222, but cannot explain the broadband emission of 980519, 990123, 990510 and 000926. 
Therefore our findings are not consistent with the "pure" winds expected from Wolf-Rayet stars 
(Chevalier \& Li 1999) and are in accord with the results of Ramirez-Ruiz \etal (2001), who have 
shown that the interaction between such winds and the circumstellar medium leads to the formation 
of a quasi-homogeneous shell extending from $\sim 10^{16}$ cm to $\simg 10^{18}$ cm.

 Regarding the microphysics of shocks, the results presented in Figure 2 indicate that 
$i)$ the total electron energy is close to equipartition with protons, 
$ii)$ the magnetic field strength, $B \propto \epsmag^{1/2}$, is within two orders of magnitude 
 of the equipartition value, and 
$iii)$ the power-law distribution of shock-accelerated electrons does not have an universal index 
 $p$, with $1.4 < p < 2.8$. 
Given that the post jet-break emission falls-off as $t^{-p}$, the shallow decays of the optical 
emission of 010222 and radio light-curves of 991208, 991216 and 000301c require models with 
$p \sim 1.5$. Consistency between such hard electron distributions and the observed optical 
spectral slopes implies that, for these four afterglows, the cooling frequency $\nu_c$ was below 
the optical domain.

 In our modelling, the passage of a high energy spectral break $\nu_*$ yields most of the steepening
of the optical decay observed in the afterglows 991208, 991216 and 000301c. This allows us to 
determine the fractional electron energy up to $\gamma_*$ corresponding to $\nu_*$. We find this 
fraction to be in the range 1/3--2/3 (Figure 2), close to equipartition, which provides a natural
reason for the existence of the break at $\gamma_*$. We note that hard electron distributions and 
equipartition electron energies find mutual consistency in the shock acceleration treatment of 
Malkov (1999). 

 Assuming an uniform jet, the jet energies and apertures inferred here and the observed $\gamma$-ray 
fluences require an efficiency of the $\gamma$-ray mechanism in dissipating the jet kinetic energy 
and radiating it in the 20 keV--1 MeV range which exceeds $\sim$ 50\%. However if the $\gamma$-ray 
emission arises from bright patches caused by angular fluctuations (Kumar \& Piran 2000) of the 
kinetic energy on the jet surface, the actual efficiency can be much smaller, closer to that obtained 
from numerical modelling of internal shocks (Spada, Panaitescu \& M\'esz\'aros 2000). These bright 
spots could also induce short timescale fluctuations in the early $X$-ray and optical afterglow 
emission, before these fluctuations disperse, and may have already been observed in the optical 
light-curve of 000301c.

\section{Conclusions}

 The findings reported above should shed some light on the GRB progenitors. Homogeneous media of 
low density, below $10^{-2} \cm3$, as found for the afterglows 990123 and 980703, indicate a 
galactic halo or a hot component of the interstellar medium. External particle densities in the 
$0.1-1\cm3$ range, as found for 980519 and 990510, are characteristic for the interstellar medium, 
while larger values of $\sim 10 \cm3$, as determined for the afterglows 991208, 000301c, 000926, 
and 010222, are typical for diffuse hydrogen clouds. Even larger particle densities, above $100 
\cm3$, would be consistent with the undisturbed, dense molecular clouds expected in the currently 
popular collapsar model (Woosley 1993, Paczy\'nski 1998, MacFadyen \& Woosley 1999) for GRB 
progenitors. Scalo \& Wheeler (2001) have emphasized that the winds and supernovae occurring in a 
cluster of massive stars create 10 pc--1 kpc superbubbles whose density can be as low as $10^{-3}
\cm3$. Furthermore they argued that variations in the cluster age and the density of the giant 
molecular cloud into which the superbubble expands, as well as the interaction with the winds 
from other clusters, may yield circumburst medium densities spanning a few orders of magnitude. 

 The relativistic kinetic energies of the eight GRBs analyzed here show a remarkably narrow 
distribution, with mean value of $\sim 3\times 10^{50}$ erg and dispersion of $\sim 10^{50}$ erg. 
On the other hand, the distributions of the $\gamma$-ray energy output (assuming uniform jets) 
and of the jet initial aperture are significantly broader, with a width of one order of magnitude.

 That half of the afterglows analyzed here are narrower than $\sim 3 \deg$ (Figure 2), indicates 
that the distribution of initial jet angles is dominated by very narrow jets.\footnote{ 
  Admittedly, our criterion for selecting the eight cases analyzed here excludes some well 
  observed afterglows for which an optical light-curve break was not seen. Such afterglows 
  should arise from wider jets, most likely with $\theta_0 \simg 20\deg$.}
Taking into account that star formation rate at redshift $z \sim 1$ was an order of magnitude 
larger than at present (Madau, Pozzetti \& Dickinson 1998), the rate of collimated GRBs is 
$\simg 10^3$ times smaller than that of supernovae. As pointed out by Paczy\'nski (2001), if the 
fraction of core collapse supernovae associated with GRBs is at $z \sim 1$ the same as in the nearby 
Universe then, within 100 Mpc, there should be $\siml 10$ GRB remnants that are sufficiently bright 
and large after several years to allow the resolution of their non-spherical structure with VLBA. 
These remnants will have typical velocities of $0.1\,c$, therefore the expansion of the nearest 
remnants, over several years, could be detectable.

 From the afterglow parameters presented in Figure 2, one can show that the jet Lorentz factor 
at the end of the GRB phase is $160 \pm 70$. For the typical jet energy given above, this implies 
that the baryonic mass ejected at ultra-relativistic speeds by the GRB explosion is about 
$10^{-6} M_\odot$. If the GRB progenitor is a $10\, M_\odot$ star, the largest jet apertures 
($\sim 10\deg$) and the baryonic loads determined here imply that the region of the star through 
which the jet propagates is evacuated up to 1 part in $10^5$.

 Our calculations led the surprising result that the electron energy distribution index varies 
significantly among GRB afterglows, suggesting that the details of the acceleration of particles 
at ultra-relativistic shocks are not universal. We also find that the energy imparted to electrons
at relativistic shocks should be close to equipartition with protons.

\acknowledgments{AP acknowledges the supported received from Princeton University through the 
       Lyman Spitzer, Jr. fellowship. PK was supported in part by the Ambros Monell foundation
       and NSF grant Phy-0070928.  We thank Bohdan Paczy\'nski, John Scalo and Craig Wheeler
       for helpful comments.}

\begin{figure}
\centerline{\psfig{figure=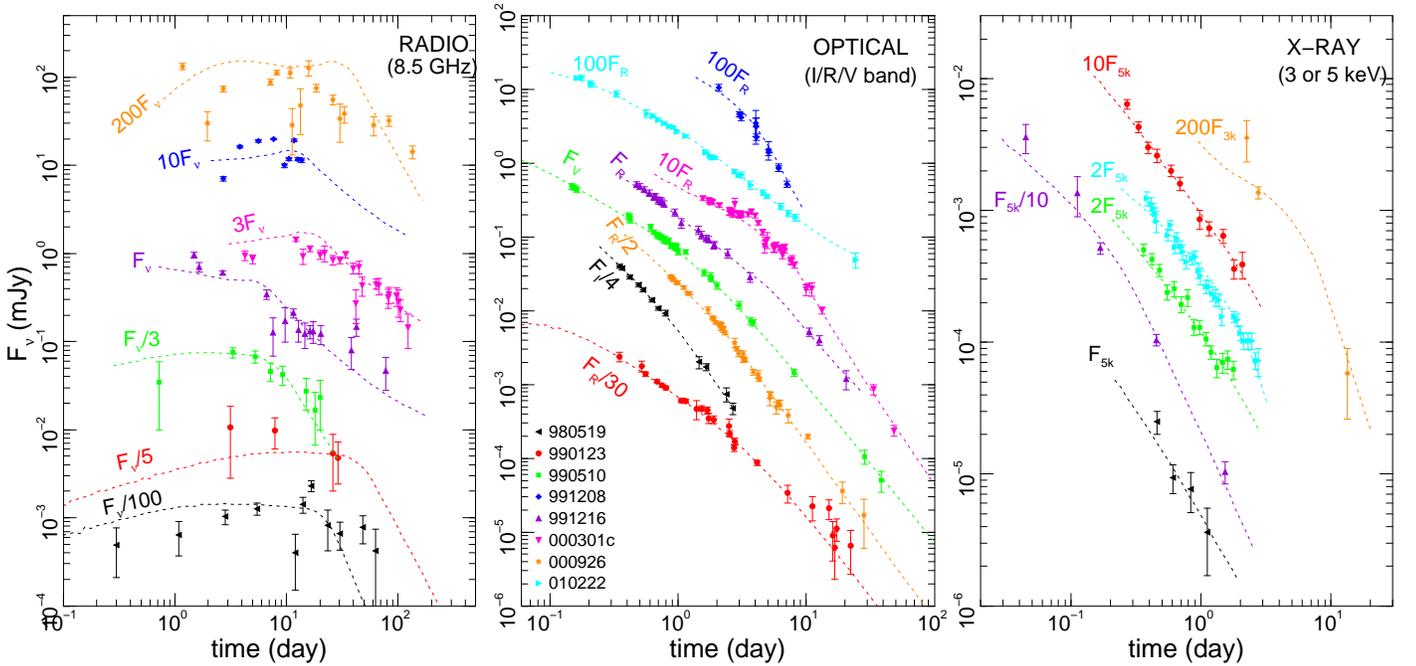}}
\figcaption{
 Radio, optical and $X$-ray emission and model light-curves for the GRB afterglows 980519, 990123, 
 990510, 991208, 991216, 000301c, 000926, and 010222 (legend of middle graph applies
 to all panels). 
 The numerical light-curves have been obtained by minimization of $\chi^2$ between model emission 
 and the radio, millimeter, sub-millimeter, near infrared, optical, and $X$-ray data (only a part 
 of the used data is shown in this figure). The parameters of each model are given in {\it Figure 2}. 
 Optical data has been corrected for Galactic dust extinction. 
 The spread around the model curves exhibited by the radio emission of 980519, 991208, 991216,
 000301c and 000926 can be explained by fluctuations due to scatterings by the inhomogeneities in 
 the Galactic interstellar medium (Goodman 1997). 
 Fluxes have been multiplied by the indicated factors, for clarity.  } 
\end{figure}

\begin{figure}
\centerline{\psfig{figure=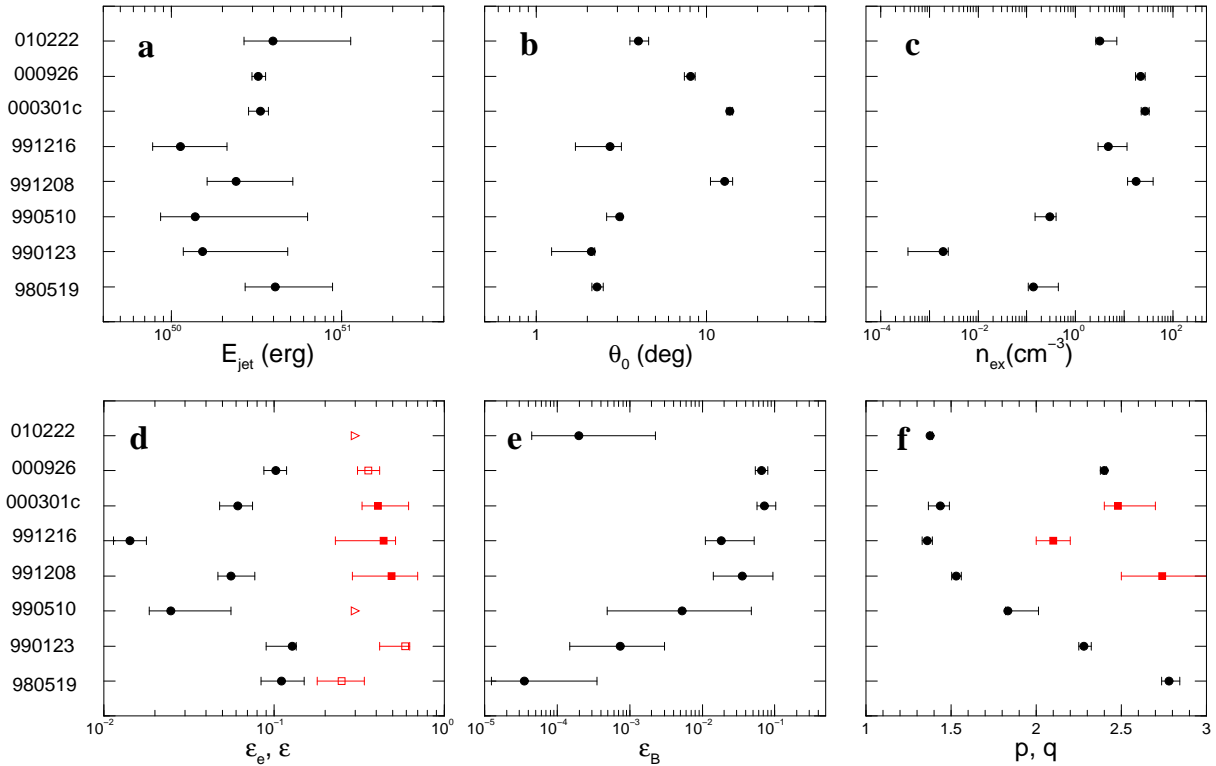,width=16cm}}
\figcaption{
 Best fit model parameters and their 90\% confidence intervals for the afterglows shown 
 in {\it Figure 1}. 
 {\bf a}: jet initial energy $E_0$, 
 {\bf b}: initial half-angle $\theta_0$,
 {\bf c}: external medium density $n$, 
 {\bf d}: parameter $\epsel$ for minimum injected electron energy (filled circles), 
          fractional electron energy $\epsilon$ up to the $\gamma_*$-break (filled squares,
          triangles indicating lower limits), and the total electron energy $\frac{p-1}{p-2}\,
          \epsel$ if the electron distribution extends to infinity (open squares), 
 {\bf e}: magnetic field parameter $\epsmag$, 
 {\bf f}: index $p$ of the power-law injected electron distribution (filled circles), and
          index $q$ above the $\gamma_*$-break (filled squares). 
  For 010222 we find an unusually small $\epsel = 0.2^{+2.1}_{-0.1} \times 10^{-3}$. } 
\end{figure}


\begin{references}

\reference{} Chevalier, R. \& Li, Z. 1999, ApJ, 520, L29
\reference{} Fenimore, E., Epstein, R. \& Ho, C. 1993, A\&AS, 97, 59
\reference{} Fishman, G. \& Meegan, C. 1995, ARA\&A, 33, 415
\reference{} Frail, D. \etal 2001, ApJL, submitted (astro-ph/0102282)
\reference{} Goodman, J. 1997, New Astronomy, 2, 449
\reference{} Kumar, P. \& Piran, T. 2000, ApJ, 535, 152
\reference{} Kumar, P. \& Panaitescu, A. 2000, ApJ, 541, L9
\reference{} MacFadyen, A., Woosley, S. \& Heger, A. 2001, ApJ, 550, 410
\reference{} MacFadyen, A., \& Woosley, S. 1999, ApJ, 524, 262
\reference{} Madau, P., Pozzetti, L. \& Dickinson, M. 1998, ApJ, 498, 106
\reference{} Malkov, M. 1999, ApJ, 511, L51
\reference{} M\'esz\'aros, P. \& Rees, M.J. 1997, ApJ, 476, 232
\reference{} M\'esz\'aros, P., Rees, M.J. \& Wijers, R. 1999, New Astronomy, 4, 303
\reference{} Paczy\'nski, B. 1998, ApJ, 494, L45
\reference{} Paczy\'nski, B. 2001, Acta Astronomica, 51, 1
\reference{} Panaitescu, A. \& Kumar, P. 2000, ApJ, 543, 66
\reference{} Panaitescu, A. \& Kumar, P. 2001, ApJ, 554, 667
\reference{} Piran, T. 1999, Phys. Rep., 314, 575
\reference{} Piro, L. 2000, Proceedings of "X-ray Astronomy '99: Stellar Endpoints, AGN 
             and the Diffuse X-ray Background", ed. N. White, Bologna (astro-ph/0001436)
\reference{} Ramirez-Ruiz, E., Dray, L., Madau, P. \& Tout, C. 2001, MNRAS, submitted
              (astro-ph/0012396)
\reference{} Rees, M.J. \& M\'esz\'aros, P. 1992, MNRAS, 258, 41p
\reference{} Rees, M.J. \& M\'esz\'aros, P. 1994, ApJ, 430, L93
\reference{} Rhoads, J. 1999, ApJ, 525, 737
\reference{} Scalo, J. \& Wheeler, J. 2001, ApJ, submitted (astro-ph/0105369)
\reference{} Spada, M., Panaitescu, A. \& M\'esz\'aros, P. 2000, ApJ, 537, 824
\reference{} Wheeler, J.C. 2000, "The Largest Explosions since the Big Bang: 
              Supernovae and GRBs", eds. M. Livio, K. Sahu \& N. Panagia, 
              Cambridge: Cambridge University Press (astro-ph/9909096)
\reference{} Woosley, S. 1993, ApJ, 405, 273

\end{references}
\end{document}